\documentclass[a4paper]{article}
\usepackage{INTERSPEECH2020}

\usepackage[american]{babel}
\usepackage{numprint}

\usepackage{amsmath,amssymb,amsfonts}
\usepackage{cite}
\usepackage{hyperref}
\usepackage{xspace}
\usepackage[linesnumbered,ruled,vlined]{algorithm2e}
\usepackage{subcaption}
\usepackage{enumitem}
\usepackage{todonotes}
\usepackage{bbm}

\newcommand{\Guesser}{Guesser\xspace}
\newcommand{\guesser}{guesser\xspace}
\newcommand{\Wordgen}{Enquirer\xspace}
\newcommand{\wordgen}{enquirer\xspace}
\newcommand{\guestemb}{voice print\xspace}
\newcommand{\guestembs}{voice prints\xspace}
\newcommand{\SRlong}{Interactive Speaker Recognition\xspace}
\newcommand{\SR}{ISR\xspace}

\title{A Machine of Few Words \\Interactive Speaker Recognition with Reinforcement Learning}
\name{Mathieu Seurin$^1$, Florian Strub$^2$, Philippe Preux$^1$, Olivier Pietquin$^3$}
\address{
  $^1$Université de Lille, CNRS, Inria - France\\
  $^2$Deepmind - France
  $^3$Google Research, Brain Team - France}
\email{Contact author : mathieu.seurin@inria.fr}

\begin{document}

\maketitle
% The total length of the abstract is limited to 200 words. The abstract included in your paper and the one you enter during web-based submission must be identical. Avoid non-ASCII characters or symbols as they may not display correctly in the abstract book.
\begin{abstract}
Speaker recognition is a well known and studied task in the speech processing domain. It has many applications, either for security or speaker adaptation of personal devices. In this paper, we present a new paradigm for automatic speaker recognition that we call Interactive Speaker Recognition (ISR). In this paradigm, the recognition system aims to incrementally build a representation of the speakers by requesting personalized utterances to be spoken in contrast to the standard text-dependent or text-independent schemes. To do so, we cast the speaker recognition task into a sequential decision-making problem that we solve with Reinforcement Learning. Using a standard dataset, we show that our method achieves excellent performance while using little speech signal amounts. This method could also be applied as an utterance selection mechanism for building speech synthesis systems. 
\end{abstract}

\noindent\textbf{Index Terms}: active speaker recognition, reinforcement learning, deep learning, iterative representation learning

\section{Introduction}

\textit{"Good words are worth much and cost little."} - George Herbert\\ In many speech-based applications, the construction of a speaker representation is required \cite{irum2019speaker}. Automatic Speaker Recognition (ASR) and Speech Synthesis are some examples that have recently made steady progress by leveraging large amounts of data and neural networks. Text-To-Speech (TTS) convincingly encompasses someone's voice \cite{shen2018natural}, and modern Speaker Recognition systems identify a speaker \cite{irum2019speaker} among thousands of possible candidates with high accuracy.
Speaker recognition systems are trained to extract speaker-specific features from speech signals, and during evaluation, test speaker utterances are compared with the already existing utterances. However, dozen of test recordings are necessary, limiting usage when interacting with humans.
When identifying a speaker or trying to create a convincing TTS system, only some key features might be necessary, such as certain inflexions or speech mannerisms. 
In this paper, we build a speaker recognition system that can identify a speaker by using a limited and personalized number of words. 
Instead of relying on full test utterance across all individuals, we interact with the speakers to iteratively select the most discriminative words.  

Some pronunciation might be typical of certain speakers. For example, the phoneme 'r' might be pronounced differently depending on your accent. Thus starting with general phoneme and refining based on the utterances received could result in better recognition systems.
More generally, a desirable feature of speaker recognition is to adapt its strategy to the current speaker as important features vary from person to person. 

Here we propose to envision the problem of building a representation of the speaker as a sequential decision-making problem. The system we want to develop will select words that a speaker must utter so that it can be recognized as fast as possible. Reinforcement learning (RL) \cite{sutton2018reinforcement} is a framework to solve such sequential decision-making problems. It has been used in speech-based applications such as dialog~\cite{chandramohan2010optimizing, strub2017end} but not to the problem of speaker identification (note that
~\cite{pietquin2005comparing} combines RL and phones similarity). We adapt a standard RL algorithm to interact with a speaker to maximize the identification accuracy given as little data as possible. After introducing an \SRlong (\SR) game based on the TIMIT dataset to simulate the speaker ASR interaction, we show that the RL agent builds an iterative strategy that achieves better recognition performance while querying only a few words. 

Our contributions are thus: 
\begin{enumerate}[leftmargin=*]
    \item to introduce the \SRlong as an interactive game between the SR module and a human (Sec.~\ref{sec:game}); 
    \item to formalize ISR as a Markov Decision Process~\cite{puterman2014markov} so as to solve the problem with RL (Sec.~\ref{sec:rl}); 
    \item to introduce a practical Deep RL \SR model, and train it on actual data (Sec.~\ref{sec:protocol}). 
\end{enumerate}

\noindent Finally, we test our method on the TIMIT dataset and show that \SR model successfully personalized the words it requests toward improving speaker identification, outperforming two non-interactive baselines (Sec.~\ref{sec:Exp}).

\section{\SRlong Game}
\label{sec:game}

    In this paper, we aim to design an \SRlong (\SR) module that identifies a speaker from a list of speakers only by requesting to utter a few user-specific words. To do so, we first formalize the \SR task as an interactive game involving the speaker and the \SR module. We then define the notation used to formally describe the game before detailing how we designed the \SR module.

    \subsection{Game Rules}\label{subsec:rules}
    
        To instantiate the \SR game, we first build a list of random individuals, or \textit{guests}. Each guest is characterized by a few spoken sentences (enrolment phase), which act as their signature that we call \textit{\guestemb}. In a second step, we label one of the guests as the target \textit{speaker} that we aim to identify. Hence a game is defined by $K$ guests characterized with $K$ \guestembs, and one of these guests is labeled as the speaker. 
        
        As the game starts, the $K$ \guestembs are provided to the \SR module, and it needs to identify the speaker among the guests. To do so, the \SR engine may interact with the speaker, but it can only request the speaker to utter $T$ words within a predefined vocabulary list. At each turn of the game, the \SR module asks the speaker to say a word, the speaker pronounces it, and the \SR engine updates its internal speaker representation, as detailed in \autoref{subsec:neural}, before asking the next word. Again, the \SR module may only request $T$ words. Thus, it needs to carefully choose them to correctly identify the speaker. 

    \subsection{Game notation}\label{subsec:taskdesc}

    A game is composed of a list of $K$ guests characterized by their \guestemb $\bm{g}=[g^k]^{K}_{k=1}$ where $\bm{g}$ is a subset from a larger group of registered guests $\mathcal{G}$ of size $G$, and a predefined vocabulary $\mathcal{V}$ of size $V$. The \SR module aims at building a list of words $\bm{w}=[w_t]^{T}_{t=1} \in \mathcal{V}$ to be uttered by the speaker. The uttered version of $\bm{w}$ is $\bm{x}=\{x_t\}_{t=1}^T$, where $x_t$ is the representation of word $w_t$ pronounced by the speaker. Note that, for a given $\bm{w}$,  $\bm{x}$ differs from one speaker to another. 

    \begin{figure}
        \centering
        \includegraphics[width=0.95\columnwidth]{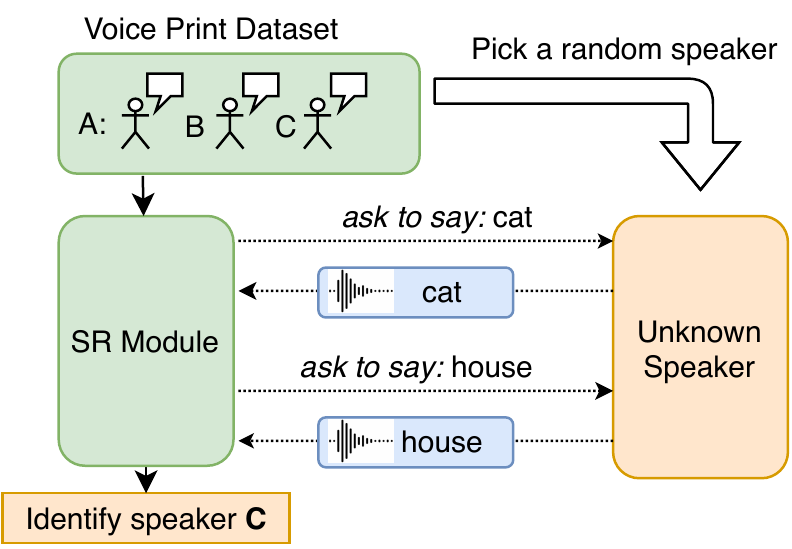}
        \caption{\SRlong game overview}
        \label{fig:game}
        \vskip -2em
    \end{figure}

    \subsection{Modelling the Speaker Recognition Module}\label{sec:modules}
    
        From a machine learning perspective, we aim to design an \SR module that actively builds an internal speaker representation to perform \guestemb classification. As further discussed in \autoref{subsec:audioprocess}, this setting differs from standard SR methods that rely on generic but often long utterances~\cite{snyder2018x}. 
        In practice, we can split this task into two sub-modules: 1) an interactive module that queries the speaker to build the representation, and 2) a module that performs the \guestemb classification. In the following, we refer to these modules as \textit{\wordgen} and \textit{\guesser}.
        
        Formally, the \guesser must retrieve the speaker in a list of $K$ guests characterized by their \guestemb $g^k \in \bm{g}$ and a sequence of words $\bm{x}$ uttered by the speaker $g^{*} \in \bm{g}$. Thus, the \guesser has to link the speaker's uttered words to the speaker's \guestemb. 
        The \wordgen must select the next word $w_{t+1} \in \mathcal{V}$ that should be pronounced by the speaker given a list of $K$ guests and %guests characterized by their \guestemb $g^i \in G$ and 
        the sequence of $t$ previously spoken words $[x_{t'}]_{t'=1}^t$. Thus, the \wordgen's goal is to pick the word that maximizes the guesser's success rate. 
        Therefore, the \SR module first queries the speaker with the \wordgen. Once the $T$ words are collected, they are forwarded to the \guesser to perform the speaker retrieval. In practice, this artificial split allows training the \guesser with vanilla supervised learning, i.e., by randomly sampling words to retrieve speakers. The \wordgen can hence be trained through reinforcement learning, as explained in the next section.

\section{Speaker Recognition as a RL Problem}\label{sec:rl}

    Reinforcement Learning addresses the problem of sequential decision making under uncertainty, where an agent interacts with the environment to maximize its cumulative reward~\cite{sutton2018reinforcement}. In this paper, we aim at maximizing the \guesser success ratio by allowing the \wordgen to interact with the speaker, which makes RL a natural fit to solve the task. In this section, we thus provide the necessary RL terminology before relating the \wordgen to the RL setting and defining the optimization protocol.

    \subsection{Markov Decision Process}\label{subsec:mdp}
    
        In RL, the environment is modeled as a Markov Decision Process (MDP), where the MDP is formally defined as a tuple $\{\mathcal{S},\mathcal{A},\mathcal{P},\mathcal{R},\gamma\}$~\cite{howard1960dynamic,puterman2014markov}.
        At each time step $t$, the agent is in a state $s_{t} \in \mathcal{S}$, where it selects an action $a_t \in \mathcal{A}$ according to its policy $\pi: \mathcal{S} \rightarrow \mathcal{A}$. The agent then moves to the state $s_{t+1}$ according to a transition kernel $\mathcal{P}$ and receives a reward $r_t=r(s_t,a_t)$ drawn from the environment's reward function $\mathcal{R}: \mathcal{S} \times \mathcal{A}$.
        In this paper, we define the \wordgen as a parametric policy $\pi_{\bm{\theta}}$ where $\bm{\theta}$ is a vector of neural network weights that will be learnt with RL. At the beginning of an episode, the initial state corresponds to the list of guests: $s_0=\{G\}$. At each time step $t$, the \wordgen picks the action $a_t$ by selecting the next word to utter $w_t$, where $w_t \sim \pi_{\bm{\theta}}(s_t)$. The speaker then pronounces the word $w_t$, which is processed to obtain $x_t$ before being appended to the state $s_{t+1} = s_t \cup \{x_t\}$. 
        After $T$ words, the state $s_T=\{\bm{g}, \bm{x}\}$ is provided to the \guesser. The \wordgen is rewarded whenever the \guesser identifies the speaker, i.e.  $r(s_t, a_t) = 0$ if $t < T$ and $r(s_T, a_T) = \mathbbm{1}_{[\text{argmax}_k p(g_k|s_T) = g^*]}$ where $\mathbbm{1}$ is the indicator function.

    \subsection{Enquirer optimization Process}\label{subsec:policy_gradient}
    
        In RL, policy search aims at learning the policy $\pi_{\bm{\theta}^*}$ that maximizes the expected return by directly optimizing the policy parameters $\bm{\theta}$. More precisely, we search to maximize the  mean value defined as $J(\bm{\theta}) = E_{\pi_{\bm{\theta}}}\big[\sum^T_{t=1}\gamma^{t-1}r(\bm{s}_t,a_t) \big]$. To do so, the policy parameters are updated in the direction of the gradient of $J(\bm{\theta})$. 
        In practice, direct approximation of $\nabla J(\bm{\theta})$ may lead to destructively large policy updates, may converge to a poor deterministic policy at early training and it has a high variance. In this paper, we thus use the recent Proximal Policy Optimization approach (PPO)~\cite{schulman2017proximal}. PPO clips the gradient estimate to have smooth policy updates, adds an entropy term to soften the policy distribution
       ~\cite{geist2019theory}, and introduce a parametric baseline to reduce the gradient variance~\cite{mnih2016asynchronous,schulman2017proximal}. 

%Following the policy gradient theorem~\cite{sutton1999policy}, this gradient can be estimated by $\nabla J(\bm{\theta}) = E_{t}\big[\pi_{\bm{\theta}}(a_t|s_t)Q^{\pi_{\bm{\theta}}}(s_t, a_t)\big]$, where $Q^{\pi_{\bm{\theta}}}$ is the state-value function defined by $Q^{\pi_{\bm{\theta}}}(s_t) = E_{\pi_{\bm{\theta}}}\big[\sum^K_{k=1}\gamma^{k-1}r(\bm{s}_{k},a_{k})| s_0=s_t, a_0=a_t \big]$. 

%%%%%%%%%%%%%%%%%%%%%

\section{Experimental Protocol}
\label{sec:protocol}

    \begin{figure*}[th]
        \centering
        \includegraphics[width=1.\textwidth]{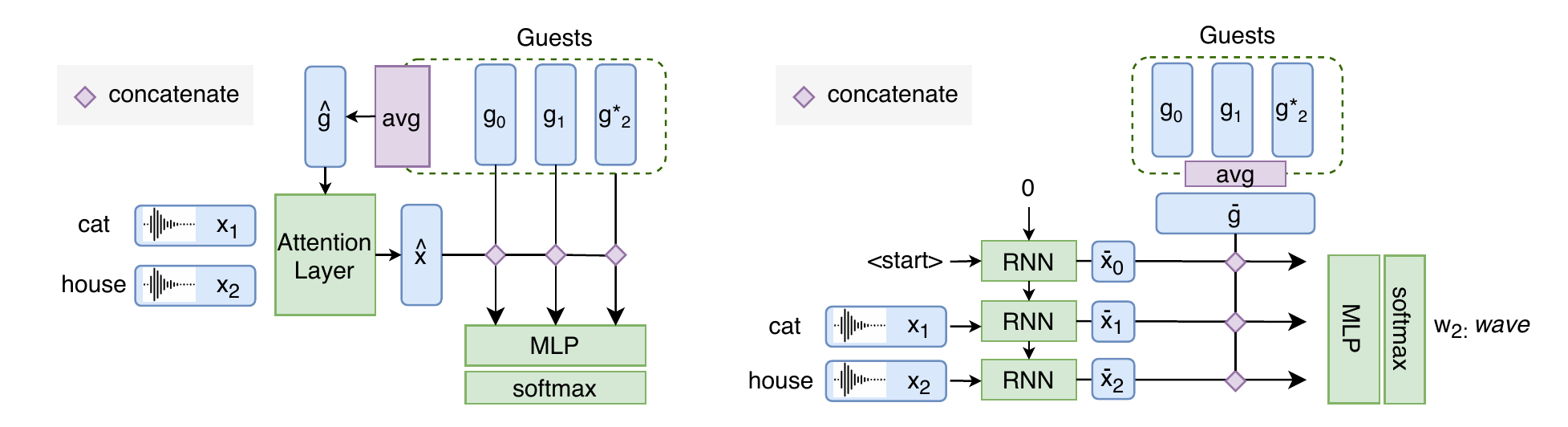}
        \caption{(Left) The \guesser must retrieve the speaker in the list of guests. (Right) The \wordgen must select the next word to utter.}
        \label{fig:models}
        \vskip -0.5em
    \end{figure*}
    
    We first detail the data we used to create the \SR game before describing the speech processing phase. Finally, we present the neural training procedure.%\footnote{The codebase and hyperparameters will be available at \url{https://github.com/Mathieu-Seurin/few_words}}
    
    \subsection{Dataset}\label{subsec:dataset}
    
        We build the \SR game using the TIMIT corpus~\cite{garofolo1992timit}. This dataset contains the voice recordings of 630 speakers with eight different accents.
        Each speaker uttered ten sentences, where two sentences are shared among all the speakers, and the eight others differ. Sentences are encoded as 16-bit, 16kHz waveforms. First, we define the \SR vocabulary by extracting the words of the two shared sentences, so the \wordgen module may always request these words whatever the target speaker. In total, we obtained twenty distinct words such as \emph{dark}, \emph{year}, \emph{carry} while dropping the uninformative specifier \emph{a}. Second, we use the eight remaining sentences to build the speakers' \guestemb.
        %['all', 'dark', 'greasy', 'had', 'in', 'she', 'suit', 'wash', 'water', 'year', 'your', 'an', 'ask', 'carry', 'dont', 'like', 'me', 'oily', 'rag', 'that', 'to']

    \subsection{Audio Processing}\label{subsec:audioprocess}
    
        Following~\cite{snyder2018x, snyder2017deep, snyder2016deep}, we first down-sample the waveform to 8kHz before extracting the Mel Frequency Cepstral Coefficient (MFCC). We use MFCCs of dimension 20 with a frame-length of 25ms, mean-normalized over a sliding window of three seconds. We then process the MFCCs features through a pretrained X-Vector network to obtain a high quality voice embedding of fixed dimension 128, where the X-Vector network is trained on augmented Switchboard~\cite{godfrey1992switchboard}, Mixer 6~\cite{chodroff2016new}, and NIST SREs~\cite{doddington2000nist}\footnote{available in kaldi library \cite{Povey_ASRU2011} at \url{http://www.kaldi-asr.org/models/m3})}. 
        To get the spoken word representation (word that the enquirer will query), we split the two shared sentences into individual words by following the TIMIT word timestamps. We then extract the X-Vector of each word $w_t$ of every speaker $k$ to obtain $x_t^k$. We compute the \guestemb by extracting the X-Vector of the eight remaining sentences before averaging them into a final vector of size 128 for each guest $g^k$.

    \subsection{Speaker Recognition Neural Modules}\label{subsec:neural}
    
        We here describe the \SR training details and illustrate the neural architectures in \autoref{fig:models}.

        \noindent \textbf{Guesser}. To model the \guesser, we first model the guest by averaging the \guestemb into a single vector $\hat{g}=\frac{1}{K}\sum{g^k}$. We then pool the X-Vectors with an attention layer conditioned on $\hat{g}$ to get the guesser embedding $\hat{x}$~\cite{bahdanau2014neural}:
        \begin{align*}
        e_t=MLP([{x}_t,\hat{g}]) \;;\;  \alpha=softmax(\bm{e}) \;;\; \hat{x}=\sum_t \alpha_t x_t,
        \end{align*}
        where [.,.] is the concatenation operator and MLP is a multilayer perceptron with one hidden layer of size 256. We concatenate the guesser embedding with the guest \guestemb before projecting them through a MLP of size 512. Finally, we use a softmax to estimate the probability $p^k$ of each guest to be the speaker, \textit{i.e.\@} $p(g^k = g^*|\bm{x}, \bm{g})=softmax\big(MLP([g^k,\hat{x}])\big)$. Both MLP have ReLU activations~\cite{nair2010rectified} with a dropout ratio of 0.5\%~\cite{srivastava2014dropout}. The guesser is trained by minimizing the cross-entropy with ADAM~\cite{kingma2014adam}, a batch size of 1024 and an initial learning rate of $3 .10^{-4}$ over \numprint{45}k games with five random guests. 

        \vspace{0.5em}
        \noindent \textbf{Enquirer}. To model the \wordgen, we first represent the pseudo-sequence of words by feeding the X-Vectors into a bidirectional LSTM~\cite{hochreiter1997long} to get the word hidden state $\bar{x}_t$ of dimension 2*128. Note that we use a start token for the first iteration. In parallel, we average the \guestemb into a single vector $\bar{g}=\frac{1}{K}\sum{g^k}$ to get the guest context. We then concatenate the word hidden state and the guest context before processing them through a one-hidden-layer MLP of size 256 with ReLU. Finally, a softmax activation estimates the probability of requesting the speaker to utter $w_{t+1}$ as the next word: $p(w_{t+1}|x_t, \cdots, x_1,\bm{g})=softmax\big(MLP([\bar{x}_t,\bar{g}])\big)$. %We further discuss the optimization process in Section~\ref{sec:rl}. 
        The \wordgen is trained by maximizing the reward encoded as the the \guesser success ratio with PPO~\cite{schulman2017proximal}. We use the ADAM optimizer~\cite{kingma2014adam} with a learning rate of 5e-3 and gradient clipping of 1~\cite{pascanu2013difficulty}. We performed 80k episodes of length $T=3$ steps and $K=5$ random guests. When applying PPO, we use an entropy coefficient of 0.01, a PPO clipping of 0.2, a discount factor of 0.9, an advantage coefficient of 0.95, and we apply four training batches of size 512 every 1024 transitions.

%%%%%%%%%%%%%%%%%%%%%

\section{Experiments}

We run all experiments over five seeds, and report the mean and one-standard deviation when not specified otherwise.% Again, the default setting consists in requesting $T=3$ words to identify the speaker among $K=5$ guests.

\label{sec:Exp}
    \vspace{-2em}
    \begin{figure*}[th]
        \begin{subfigure}[c]{0.33\linewidth}
            \centering
            \includegraphics[width=\linewidth]{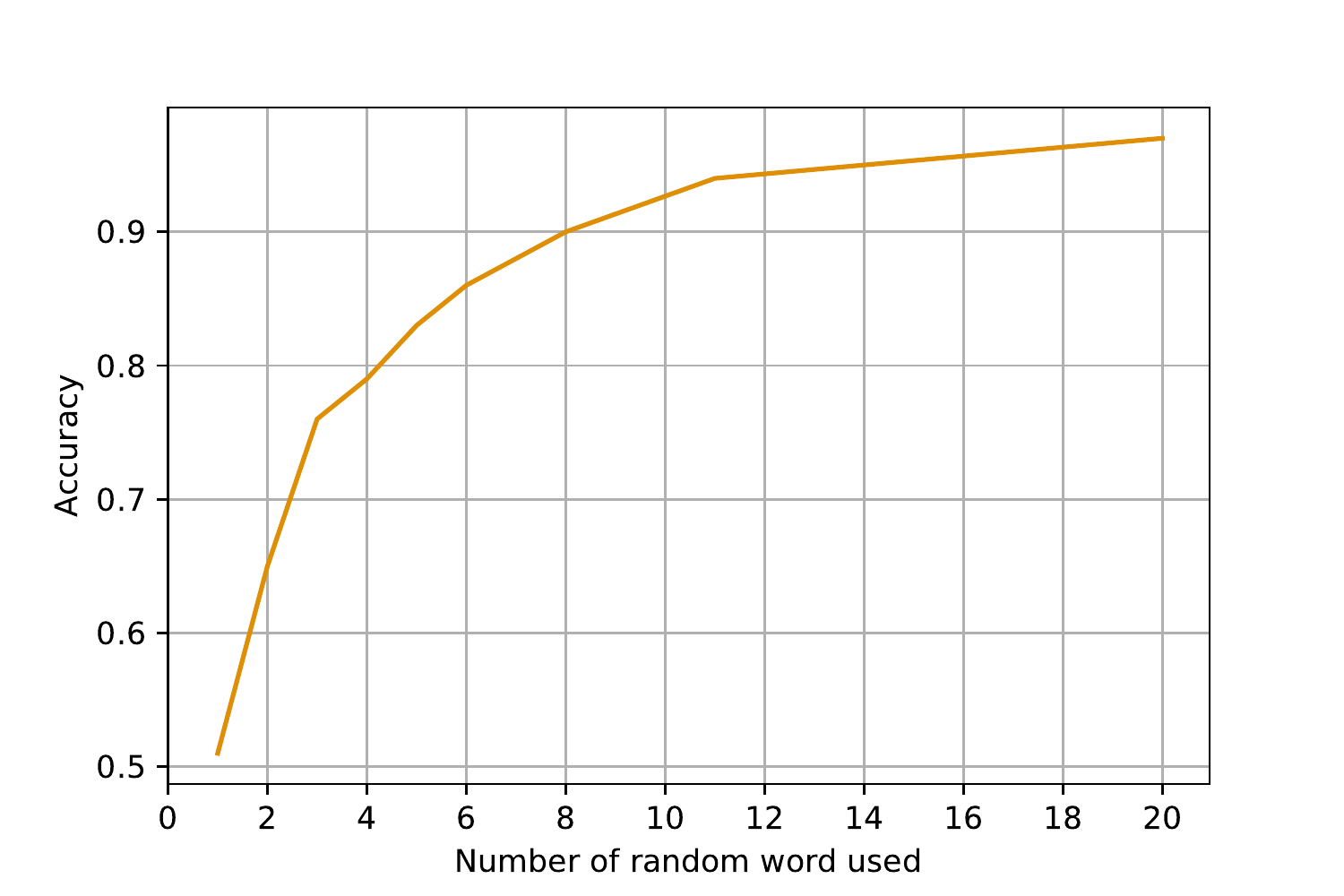}
            \caption{}\label{fig:guesser_acc_word}
        \end{subfigure}
        \hfill%
        \begin{subfigure}[c]{0.33\linewidth}
            \includegraphics[width=\linewidth]{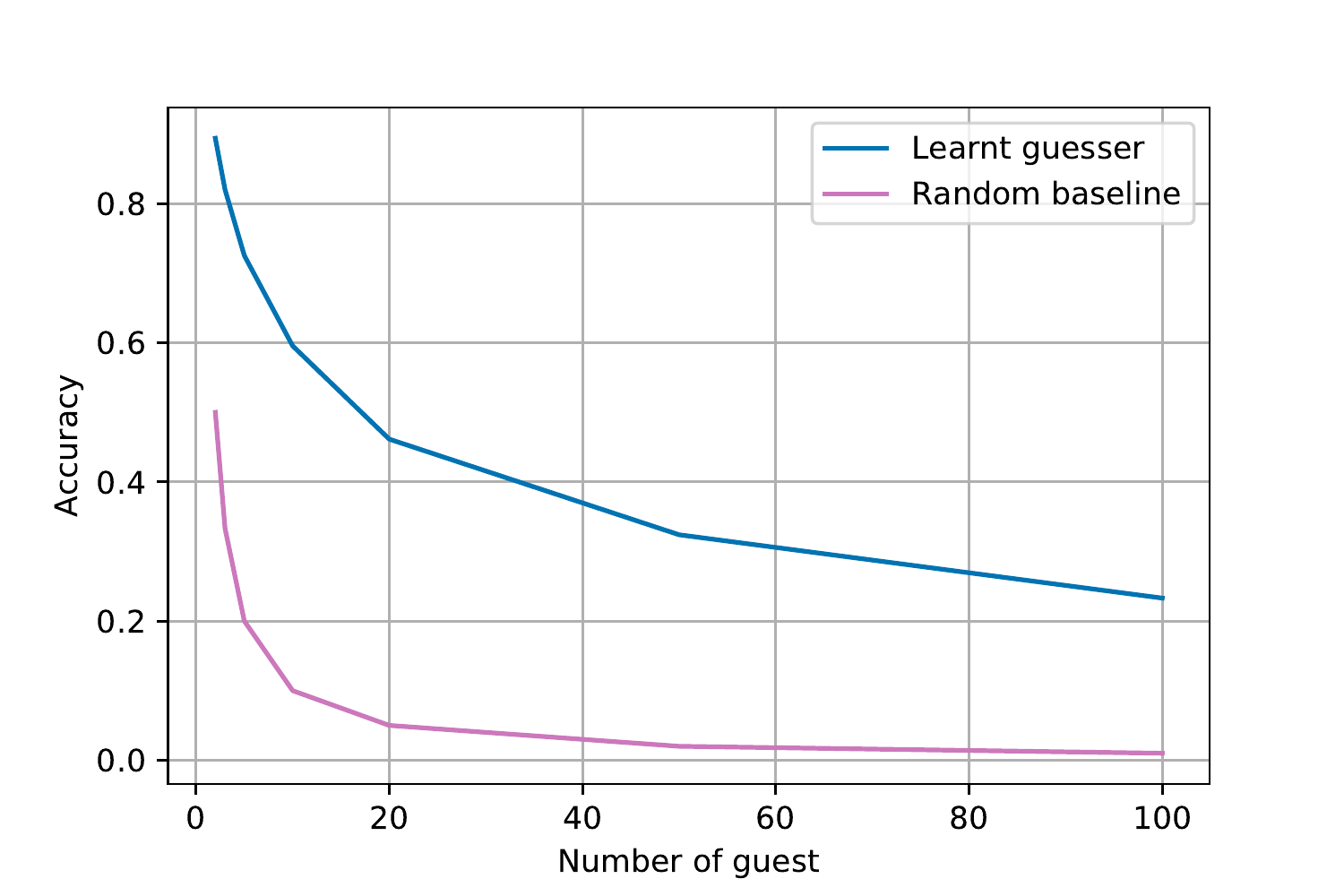}
            \caption{}\label{fig:guesser_acc_guest}
        \end{subfigure}
        \hfill%
        \begin{subfigure}[c]{0.33\linewidth}
            \centering
            \includegraphics[width=\linewidth]{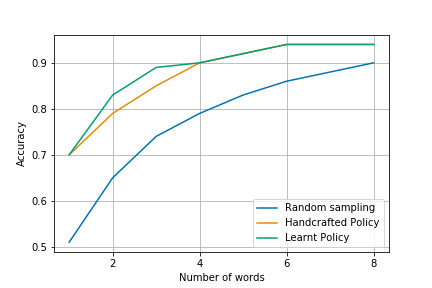}
            \caption{}\label{fig:enquirer_num_words}
        \end{subfigure}
        \vskip -0.8em
        \caption{(a-b) \Guesser test accuracy, respectively varying the number of words (resp. guests) being used (c) \Wordgen test accuracy varying the number of queried words. The RL \wordgen outperforms the heuristic baseline when selecting a low number of words.}
        \vskip -0.5em
    \end{figure*}

\subsection{Guesser Evaluation}

    In this section, we evaluate the guesser accuracy in different settings. As mentioned, we opt to request $T=3$ words to identify the speaker among $K=5$ guests. In this default setting, a random policy has a success ratio of $20\%$, whereas the neural model reaches $74.1\%\pm 0.2$ on the test set. 
    As the \guesser is trained on random words, these scores may be seen as an \SR lower-bound for the \wordgen, which would later refine the word selection toward improving the guesser success ratio.
    Thus, this setting shows an excellent ratio between task difficulty and \guesser initial success, allowing to train the \wordgen with a relatively dense reward signal. 

    \vspace{0.5em}
    \noindent \textbf{Word Sweep}. We assess the \guesser quality to successfully perform speaker recognition when increasing the number of words $T$ in \autoref{fig:guesser_acc_word}. We observe that a single word only gives $50\%$ speaker retrieval, but the accuracy keeps improving when requesting more words. Noticeably, collecting the full vocabulary only scores up to 97\% accuracy.
    
    \vspace{0.5em}
    \noindent \textbf{Guest Sweep}. We report the impact of the number of guests $K$ in \autoref{fig:guesser_acc_guest}. The guesser accuracy quickly collapses when increasing the number of guests with $K=50$ having a $46\%$ success ratio. As the number of words remains small, the \guesser experiences increasing difficulty in discriminating the guests. One way to address this problem would be to use a Probabilistic Linear Discriminant Analysis (PLDA) \cite{ioffe2006probabilistic} to enforce a discriminative space and explicitly separate the guests based on their class.

    \subsection{\Wordgen Evaluation}\label{subsec:enquirer_eval}
    
        \textbf{Model}. As previously mentioned, the \wordgen aims to find the best sequence of words $\bm{w}$ that maximizes the \guesser accuracy by interacting with the speaker. At each time step, we thus select the word with the highest probability $p(w_{t+1}|x_t, \cdots, x_1,\bm{g})$ according to the policy without replacement, i.e., the model never requests the same word twice. %We finally report the guesser scores given $\bm{w}$.
        
        \vspace{0.5em}
        \noindent \textbf{Baseline}. We compare our approach to two baselines: a random policy, and a heuristic policy. As the name suggests, the random baseline picks $T$ random words without replacement. %On the other hand, the deterministic baseline always selects the same $T$ words independently of the guests. 
        To obtain a strong baseline, we pre-select words by taking advantage of the \guesser model, where we value a sequence of words by computing the guesser accuracy over $\eta=20000$  games. Optimally, we want to iterate over every tuple of words to retrieve the optimal set; yet, it is computationally intractable as it requires $\eta*{V\choose T}$ estimations. Therefore, we opt for a heuristic sampling mechanisms. We curated a list of the most discriminant words (words that increase globally the recognition scores) and sample among those instead of the whole list.
        
        %we opt for a heuristic policy that only requires $\eta*T*V$ guesser estimations. Given an empty tuple $\bm{\omega}_1$, we look for the word $w^{*}_{t} \in \mathcal{V}$, such as $\bm{\omega}_t \cup \{w^{*}_{t}\}$ returns the highest guesser accuracy. We repeat this operation $T$ times to obtain the heuristic deterministic policy characterized by $\bm{\omega}_T$. Note that the heuristic policy may have some variance due to the guesser estimation, but it is empirically negligible.
        
\begin{figure}[t!]
            \vskip -1.8em
            \centering
            \includegraphics[width=0.95\columnwidth]{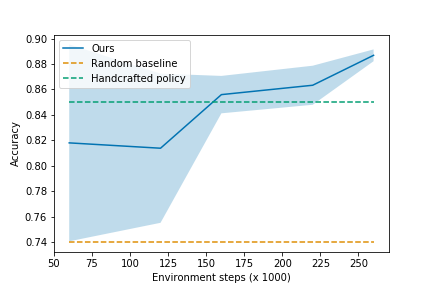}
            \caption{\Wordgen test accuracy averaged over 3 random seeds}
            \label{fig:enquirer_fixed_compa}    
            \vskip -1.5em
        \end{figure}

        \vspace{0.5em}
        \noindent \textbf{Results}. In our default setting, the random baseline reaches $74.1\%\pm 0.2$ speaker identification, and the heuristic baseline scores up to $85.1\%$. The RL \wordgen obtains up to $88.6\%\pm 0.5$, showing that it successfully leverages the guests' \guestembs to refine its policy. We show the RL training in \autoref{fig:enquirer_fixed_compa}. At early training, we observe that the \SR module still has high variance, and may behave randomly. However, RL \wordgen steadily improves upon training, and it consistently outperforms the heuristic baseline.

        \vspace{0.5em}
        \noindent \textbf{Word Diversity}. To verify whether the \wordgen adapts its policy to the guests, we generate a game for every speaker in the test set, and collect the requested words. We then compute the overlap $\Omega$ between the tuple of words by estimating the averaged Jaccard-index~\cite{jaccard1901distribution} of every pair of speakers as follow:
        \begin{align*}
        \Omega = \frac{1}{\sum^{N-1}_{n} n}\sum^{N-1}_{i=1} \sum^{N}_{j=i} J(\bm{w}^i, \bm{w}^j)\;;\;  \text{where J}(A,B) =  \frac{A \cap B}{A \cup B}
        \end{align*}
        where $N$ is the number of speakers in the test set and $\bm{w}^i$ is the word tuple of game $i$. Intuitively, the lower this number, the more diverse the policy, e.g, the deterministic policy have a Jaccard-index of 1. In the default setting, the random policy has an index of 0.14 while the RL agent has an index of 0.65. Thus, the requested words are indeed diverse.

        \vspace{0.5em}
        \noindent \textbf{Requesting Additional Words} We here study the impact of increasing the number of words $T$ requested by the \wordgen (see  \autoref{fig:enquirer_num_words} for results). First, we observe that the \SR module manages to outperform the heuristic policy when requesting two to four words, showing that the interaction with the speaker is beneficial in the low data regime. This effect unsurprisingly diminishes when increasing the number of words. However, we noticed that the \wordgen always outputs the same words when $t=1$. It suggests that the model faces some difficulties contextualizing the guests' \guestemb before listening to the first speaker utterance. We assume that more advanced multimodal architecture, e.g., multimodal transformers~\cite{lu2019vilbert,tan2019lxmert}, may ease representation learning, further improving the \SR agent.

\section{Conclusions and Future Directions}

In this paper, we introduced the \SRlong paradigm as an interactive game to improve speaker recognition accuracy while querying only a few words. We formalize it as a Markov Decision Process and train a neural model using Reinforcement Learning. We showed empirically that the \SR model successfully personalizes the words it requests to improve speaker identification, outperforming two non-interactive baselines. 
Future directions can include : scaling to bigger datasets \cite{Nagrani17, Chung18b}, scaling up vocabulary size \cite{dulac2015deep, he2016deep, seurin2019m}
Our protocol may go beyond speaker recognition. The model can be adapted to select speech segments in the context of Text-To-Speech training. Interactive querying may also prevent malicious voice generator usage by asking complex words to the generator in a speaker verification setting.

\section{Acknowledgements}

We would like to thank Corentin Tallec for his helpful recommendations and the SequeL-Team for being a stimulating environment. We acknowledge the following agencies for research funding and computing support: Project BabyRobot (H2020-ICT-24-2015, grant agreement no.687831), and CPER Nord-Pas de Calais/FEDER DATA Advanced data science and technologies 2015-2020

%The ISCA Board would like to thank the organizing committees of the past INTERSPEECH conferences for their help and for kindly providing the template files. \\
%Note to authors: Authors should not use logos in acknowledgement section; rather authors should acknowledge corporations by naming them only.

\bibliographystyle{IEEEtran}

\bibliography{mybib}

\end{document}